\documentclass[useAMS,usenatbib]{mn2e}
\usepackage[dvips]{graphicx} 

\title[Velocities
with the Sunyaev-Zeldovich effects: scaling relations and
systematics]{Measuring cluster peculiar velocities 
with the Sunyaev-Zeldovich effects: scaling relations and systematics}
\author[Diaferio et al.]{A. Diaferio,$^1$\thanks{E-mail: diaferio@ph.unito.it (AD); 
borgani@ts.astro.it (SB); lauro.moscardini@unibo.it (LM); giuseppe@to.astro.it (GM); 
kdolag@pd.astro.it (KD); volker@mpa-garching.mpg.de (VS);
tormen@pd.astro.it (GT); tornatore@ts.astro.it (LT); tozzi@ts.astro.it (PT)} 
S. Borgani,$^{2,3\star}$ L. Moscardini,$^{4\star}$ G. Murante,$^{5\star}$ K. Dolag,$^{6\star}$   
\newauthor V. Springel,$^{7\star}$ G. Tormen,$^{6\star}$ L. Tornatore,$^{2\star}$ P. Tozzi$^{8\star}$\\
$^1$Dipartimento di Fisica Generale ``Amedeo Avogadro'', Universit\`a degli Studi di Torino, Via P. Giuria 1, I-10125, Torino, Italy\\
$^2$Dipartimento di Astronomia, Universit\`a di Trieste, Via Tiepolo 11, I-34131, Trieste, Italy\\
$^3$INFN -- National Institute for Nuclear Physics, Trieste, Italy\\
$^4$Dipartimento di Astronomia, Universit\`a di Bologna, via Ranzani 1, I-40127, Bologna, Italy\\
$^5$INAF, Osservatorio Astronomico di Torino, Strada Osservatorio 20, I-10025, Pino Torinese, Italy\\
$^6$Dipartimento di Astronomia, Universit\`a di Padova, vicolo dell'Osservatorio 2, I-35122, Padova, Italy\\
$^7$Max-Planck-Institut f\"ur Astrophysik, Karl-Schwarzschildstr. 1, Garching bei M\"unchen, Germany\\
$^8$INAF, Osservatorio Astronomico di Trieste, via Tiepolo 11, I-34131, Trieste, Italy
}
\begin{document}
\maketitle
\begin{abstract}

The fluctuations in the Cosmic Microwave Background (CMB) intensity
due to the Sunyaev-Zeldovich (SZ) effect are the sum of a thermal and
a kinetic contribution. Separating the two components to measure the
peculiar velocity of galaxy clusters requires radio and microwave observations at
three or more frequencies, and knowledge of the temperature $T_e$ of the intracluster 
medium weighted by the electron number density.  To
quantify the systematics of this procedure, we extract a sample of 117
massive clusters at redshift $z=0$ from an $N$-body hydrodynamical
simulation, with $2\times 480^3$ particles, of a cosmological volume
192 $h^{-1}$ Mpc on a side of a flat Cold Dark Matter model with 
$\Omega_0=0.3$ and $\Omega_\Lambda=0.7$. Our simulation 
includes radiative cooling, star formation and the effect of feedback
and galactic winds from supernovae. We find that (1) our
simulated clusters reproduce the observed scaling relations between X-ray and SZ properties;
(2) bulk flows internal to the intracluster medium affect the velocity estimate by less
than 200 km~s$^{-1}$ in 93 per cent of the cases; (3) using the X-ray emission weighted temperature, as an estimate of $T_e$,
can overestimate the peculiar velocity by $20-50$ per
cent, if the microwave observations do not spatially resolve the cluster. For spatially
resolved clusters, the assumptions on the spatial distribution of the ICM,
required to separate the two SZ components, still produce 
a velocity overestimate of $10-20$ per cent, even with an unbiased measure of $T_e$.
Thanks to the large size of our cluster samples, 
these results set a robust lower limit of $\sim 200$ km~s$^{-1}$ to the systematic errors that will affect
upcoming measures of cluster peculiar velocities with the SZ effect. 

\end{abstract}
\begin{keywords}
large-scale structure of Universe -- galaxies: clusters: general -- cosmology:
miscellaneous -- methods: numerical
\end{keywords}

\section{Introduction}

The properties of the spatial distribution and velocity field of
matter in the Universe constrain the models of large-scale structure
formation. The velocity field has mostly been used to estimate
$\Omega_0$, the mass density parameter of the Universe, by means of the
clustering anisotropy in redshift space (e.g. \citealt{hamilton98};
\citealt{matsu02}), the connection between the density and the
peculiar velocity fields (\citealt{courteau01} and references
therein), or the mean relative peculiar velocity of galaxy pairs as a
function of pairwise separation \citep{feldman03}.
 
Further constraints on structure formation models can be extracted
from the evolution of the velocity field on large scales. Its measure
is possible with the detection of the kinetic SZ effect
\citep{sunyaev72}.  When the Cosmic Microwave Background (CMB) photons
are scattered by the free electrons in the intracluster medium (ICM),
the CMB radiation varies its intensity and its spectral distribution.
The thermal motion of the free electrons and their bulk velocity have
a different spectral signature on the CMB and the two contributions
can be separated with multi-frequency observations. However, the bulk
velocity contribution (the kinetic SZ effect) is, on average, an order
of magnitude smaller than the thermal contribution (the thermal SZ
effect), and leaves the CMB spectrum unchanged: these features make
the kinetic SZ effect difficult
to detect, although it has been recently shown how its spatial
correlation with the thermal SZ effect (\citealt*{diaf00}; \citealt{dasilva01}) can be used to remove the CMB contribution
efficiently \citep{forni04}.
 
The thermal effect is now becoming routinely detected and major
efforts have been done to plan several SZ cluster surveys which will
produce catalogues of tens of thousands of clusters in the coming
years (see \citealt*{carlstrom02} for a review and \citealt{schulz03} for systematics). 
The SZ effect does
not suffer from the usual redshift dimming of electromagnetic surface
brightnesses, and clusters can be detected,
in principle, as far back in time as when they formed. 

The distribution of SZ clusters on the sky can already provide
information on cosmological parameters (\citealt{mei03}, 2004) and cluster
biasing (\citealt{diaf03}; \citealt{cohn04}), although one has to separate the effects of
the ICM properties from those of the cosmological model
\citep{mosca02}.  Stronger constraints for cosmology are of course
provided by the cluster redshift distribution
(e.g. \citealt*{hold01}; \citealt*{benson02}; \citealt*{weller02}).
The accuracy of cluster redshifts determined with
photometric observations \citep{huterer04} or methods based on the
SZ information alone (\citealt{diego03}; \citealt*{zaroubi03})
may be sufficient to constrain cosmology, and 
optical or X-ray follow-up
spectroscopy, which can be particularly demanding in
terms of observation time and sensitivity, may not be needed for a 
first analysis.

If we measure the peculiar velocity of clusters, we can
reconstruct the redshift evolution of the velocity field on large scales.
Many authors have investigated the feasibility of measuring the
bulk flow on $\sim 100 h^{-1}$ Mpc scales with the kinetic SZ effect
(\citealt{haehnelt96}; \citealt{kash00}; \citealt*{aghanim01}; \citealt*{atrio04}). No other
method is currently viable for this task other than the kinetic SZ effect. 

To date, because of the weakness of the total SZ signal compared to the sensitivity
of the radio telescopes, separating the kinetic from the thermal contribution has
required the estimate of the expected SZ signal with an {\it a priori} 
knowledge of the temperature and spatial distribution of the ICM.
X-ray observations have been used to obtain this information for spatially well resolved clusters
(e.g. \citealt{holz97}; \citealt{laroque03}; \citealt{benson03}, 2004; \citealt{kitayama04}). 

This procedure can not be easily extended to clusters at high redshift where
X-ray observations can be beyond the limit of detectability and spatial resolution
can be poor.  In principle, but only for massive clusters
with hot ICM ($\ga 8$ keV), one can estimate the temperature, rather than with X-ray observations, by using 
the dependence of the spectral function of the SZ effect on the gas temperature, 
which appears when a fully relativistic treatment of the photon scatter is considered (\citealt*{point98};
\citealt*{itoh98}).
However, a recent attempt, although successful,
has provided a temperature measure with quite large uncertainties
\citep*{hansen02}.
 
Further insights into the physical properties of the ICM come from the
scaling relations between X-ray and SZ observables.
Analytic models (\citealt{dossantos02}; \citealt{mccarthy03a}), semi-analytical models (\citealt{cavaliere01};
\citealt{verde02}), and $N$-body hydrodynamical simulations (e.g. \citealt{dasilva03})
have been used to investigate the
effects of heating, cooling, and preheating on the normalization, slope and scatter of these scaling relations.

Here, we use a large $N$-body hydrodynamical simulation
\citep{borgani03}, which includes an advanced treatment of the gas dynamics,
contains a large sample of simulated massive clusters and reproduces
reasonably well the X-ray properties of real clusters, to verify (1) whether we
can reproduce the observed scaling relations between X-ray and SZ
observables; (2) whether the two basic assumptions underlying the
peculiar velocity measurement hold in the simulations; namely whether
(i) the mean gas velocity equals the dark matter bulk velocity and
the internal gas bulk flows are negligible \citep*{nagai03}, and (ii)
the ICM temperature derived from X-ray observations is a good
estimator of the ICM temperature weighted by the electron number
density. Our simulated sample contains 117 well resolved clusters and provides a 
major statistical improvement on previous analyses performed on a handful number of clusters
(e.g. \citealt{nagai03}; \citealt{hold03}).

We do not investigate the systematics due to a realistic
observational procedure, which includes cleaning the radio detection
of instrumental noise, atmospheric emission, galactic emission at low
frequencies, galactic dust and young galaxy emission at high
frequencies, radio sources, and the CMB itself. A detailed analysis
of these effects can be found, for example, in  \citet*{knox04} and \citet*{aghanim04}. Our results
will only provide a lower limit to the systematic uncertainties on the
peculiar velocity measures.

The layout of this paper is as follows: 
in Sect. \ref{sec:basics} we summarize the basic equations used to estimate the peculiar
velocity with the SZ effect; in Sect. \ref{sec:model} we describe the $N$-body simulation
and our sample of simulated clusters; Sect. \ref{sec:scaling} discusses the scaling
relation between X-ray and SZ observables and Sect. \ref{sec:assumptions} tests the 
basic assumptions on the velocity measures. In Sect. \ref{sec:velocity} we estimate
the systematic errors affecting the peculiar velocity measurements. 
Our conclusions are in Sect. \ref{sec:end}.

\section{Basics}\label{sec:basics}

Here, we describe the basic equations governing the SZ effects and their
observational analysis. The hottest cluster in our simulated sample has temperature
$\approx 7$ keV and we can safely ignore any relativistic corrections which are relevant
for clusters with an ICM temperature $\ga 8$ keV.

\subsection{Resolved Sources}

At a given frequency $\nu$, the variation in the CMB specific intensity due to the 
thermal motion of free electrons in the ICM 
is, in the limit of non-relativistic electron speeds (e.g. \citealt{rephaeli95}),

\begin{equation}
\Delta I_t(x,\btheta) = i_0 g(x) y(\btheta)\; , 
\end{equation}
where $x=h_P\nu/kT$, $k$ is the Boltzmann constant, $T=2.725$ K the present
day CMB temperature \citep{mather99}, $h_P$ the Planck constant, 
$i_0=2(kT)^3/(h_Pc)^2=1.129\cdot 10^3 (T/K)^3$ mJy arcmin$^{-2}$ = $13.34 (T/K)^3$ MJy
sr$^{-1}$, $c$ the speed of light, and  
\begin{equation}
g(x)={x^4 e^x\over (e^x-1)^2} \left(x{e^x+1\over e^x-1}-4\right)
\end{equation} 
the spectral function. The  Comptonization parameter is
\begin{equation}
y(\btheta)= \langle\tau(\btheta)\rangle_{\rm los} 
{k\langle T_e(\btheta)\rangle_{\rm los} \over m_e c^2} \; , 
\end{equation}
where $m_e$ is the electron mass,
\begin{equation}
\langle\tau(\btheta)\rangle_{\rm los} =\sigma_T \int n_e(r) {\rm d}l \; ,
\end{equation} 
\begin{equation}
\langle T_e(\btheta) \rangle_{\rm los} = {\int n_e(r)T_e(r) {\rm d}l \over \int n_e(r) {\rm d}l} \; ,
\label{eq:Telmap}
\end{equation}
and $\sigma_T$ is the Thomson cross section;
$n_e$ and $T_e$ are the electron number density and temperature, respectively. 
The integrals
are over the line of sight $l$, and $r^2=d_A^2\theta^2+l^2$, where 
$d_A$ is the angular diameter distance
to the cluster, and $\btheta$ is the angular separation vector from the cluster centre.
$\langle T_e(\btheta) \rangle_{\rm los}$ is the optical depth weighted temperature 
and it generally differs from the pressure-weighted temperature
\begin{equation}
\langle T_p(\btheta) \rangle_{\rm los} = {\int n_e(r)T_e^2(r) {\rm d}l \over \int n_e(r) T_e(r) {\rm d}l} \; .
\label{eq:Telpressure}
\end{equation}
$\langle T_p(\btheta) \rangle_{\rm los}$ can be derived from SZ observations alone by including 
the relativistic correction
in the spectral function $g(x)$ (\citealt{hansen04}; \citealt{knox04}).
In the following, we will not consider this temperature because
our simulated clusters are not massive enough to make the relativistic correction 
relevant; therefore, with the sensitivity of the telescopes available now or in the near future
at the radio and microwave wavelengths,
the pressure-weighted temperature $\langle T_p(\btheta) \rangle_{\rm los}$ 
of our simulated clusters will not be measurable. 

The bulk flows of both the ICM within the cluster and the cluster itself yield a further
variation in the CMB intensity: 

\begin{equation}
\Delta I_k(x,\btheta) = -i_0h(x)b(\btheta)\; , 
\end{equation}
\begin{equation}
h(x) = {x^4 e^x\over (e^x-1)^2}\; ,
\end{equation}   
\begin{equation}
b(\btheta) = \langle \tau(\btheta)\rangle_{\rm los} \langle\beta(\btheta)\rangle_{\rm los}\; ,
\end{equation}
\begin{equation}
\langle\beta(\btheta)\rangle_{\rm los} = {1\over c} {\int n_e(r)v_{\rm los}(r) {\rm d}l\over \int n_e(r) {\rm d}l} \; .
\label{eq:bel}
\end{equation} 

One actually observes the sum of the thermal and kinetic contributions
\begin{equation}
\Delta I(x,\btheta) = i_0[g(x) y(\btheta)  - h(x) b(\btheta)]\; . 
\end{equation}
At $\btheta=\bmath{0}$, we have
\begin{equation}
\Delta I_0(x) = i_0 \left[g(x)y_0-h(x)b_0\right]\; ,
\label{eq:DeltaI}
\end{equation}
where $y_0=\tau_0(kT_e/m_ec^2)$, $\tau_0=\langle \tau(\bmath{0})\rangle_{\rm los}$, 
$T_e=\langle T_e (\bmath{0})\rangle_{\rm los}$,
and $b_0=\tau_0\beta$, $\beta=\langle\beta(\bmath{0})\rangle_{\rm los}$.

Usually, to extract the SZ image of the cluster from the radio/microwave data, a $\beta$-model of the gas
distribution is assumed.
The shape parameters $\beta_c$ and $\theta_c$ are derived from a
fit to the X-ray surface brightness
\begin{equation}
\Delta S(\btheta) = \Delta S_0 \left(1+{\btheta^2\over \theta_c^2}\right)^{1/2-3\beta_c}\; .
\end{equation}
At a frequency $x$, the CMB intensity variation due to the SZ effects is expected to be
\begin{equation}
\Delta I(x,\btheta) = \Delta I_0(x) \left(1+{\btheta^2\over \theta_c^2}\right)^{(1-3\beta_c)/2}\; ,
\end{equation}
where $\Delta I_0(x)$ is now the only parameter provided by the
microwave observations. 
When we have a measure of $\Delta I_0(x)$ at three or more frequencies $x$, 
a fit to equation (\ref{eq:DeltaI}) provides an estimate of $y_0$ and $b_0$.
Simultaneous multi-frequency observations can be performed to minimize the
variations in the atmospheric emission. In this case
the fitting procedure to the spectral function is obviously
considerably more complicated,
because the $\Delta I_0(x)$'s are correlated \citep{benson03}.
 
By combining $y_0$ and $b_0$, one finally derives the line-of-sight peculiar velocity:
\begin{equation}
\beta = {k T_e\over m_ec^2} {b_0\over y_0}\; .
\label{eq:vel-res}
\end{equation}
Provided that the spherical $\beta-$model adequately describes the gas distribution,
when applied to real clusters, this equation has two main problems, that
we will investigate below: (1) 
the central value of the velocity $\beta=\langle\beta(\bmath{0})\rangle_{\rm los}$ 
might not necessarily coincide with the global peculiar velocity of the cluster;
(2) only the temperature averaged within the region of the size of the beam is actually available,
rather than the temperature in the cluster centre $T_e=\langle T_e (\bmath{0})\rangle_{\rm los}$. 
                                      
\subsection{Unresolved sources}\label{sec:unresolved}

If the spatial resolution of the radio telescope is not sufficient to resolve the source,
the observable quantity is the variation of the
CMB specific intensity due to the SZ effect integrated over the
beam size $\theta_b$ (we assume here an ideal
step function window)
\begin{equation}
\Delta S(x,\theta_b) = \int_{\theta_b} \Delta I(x) {\rm d}\Omega =
i_0[g(x)Y(\theta_b)  - h(x) B(\theta_b)]\; ,
\end{equation}
where
\begin{equation}
Y(\theta_b)=\int_{\theta_b} y(\btheta) {\rm d}\Omega=\langle\tau\rangle_{\theta_b} {k\langle T_e\rangle_{\theta_b}\over m_ec^2},
\end{equation}
\begin{equation}
B(\theta_b)=\int_{\theta_b} b(\btheta) {\rm d}\Omega = \langle\tau\rangle_{\theta_b} \langle\beta\rangle_{\theta_b},   
\end{equation}
\begin{equation}  
\langle\tau\rangle_{\theta_b} =\sigma_T \int_{r_A(\theta_b)} n_e(r) {\rm d}l {\rm d}\Omega = {\sigma_T \over d_A^2} \int n_e(r) {\rm d}^3r, 
\end{equation} 
\begin{equation}
\langle T_e \rangle_{\theta_b} = {\int_{r_A(\theta_b)} n_e(r)T_e(r) {\rm d}^3r\over \int_{r_A(\theta_b)} n_e(r) {\rm d}^3r}, 
\label{eq:Tel}
\end{equation}
and
\begin{equation}
\langle\beta\rangle_{\theta_b} = {1\over c} {\int_{r_A(\theta_b)} n_e(r)v_{\rm los}(r) 
{\rm d}^3r \over \int_{r_A(\theta_b)} n_e(r) {\rm d}^3r}, 
\end{equation}  
where $r_A(\theta_b)$ is the radius of the solid angle circle in physical units,
$dl$ is the proper coordinate element along the line of sight,
and $d^3r=d_A^2 dl d\Omega$ is the proper volume element.
With radio/microwave observations
at three or more frequencies, we can determine $Y(\theta_b)$ and
$B(\theta_b)$ as
free fit parameters. The peculiar velocity is thus
\begin{equation}
\langle\beta\rangle_{\theta_b} = {k\langle T_e\rangle_{\theta_b}\over m_ec^2}
{B(\theta_b)\over Y(\theta_b)}\; .
\label{eq:vel-unres}
\end{equation}   

\section{The Model}\label{sec:model}

\subsection{The Simulated Cluster Sample}

\citet{borgani03} describe in detail the simulation we use
here. Briefly, we simulate a cubic volume, 192 $h^{-1}$ Mpc on a side,
of a flat $\Lambda$CDM universe, with matter density $\Omega_0=0.3$,
Hubble constant $H_0=100 h$ km s$^{-1}$ Mpc$^{-1}$, $h=0.7$, baryon
density $\Omega_{\rm bar}=0.02 h^{-2}$ and power spectrum
normalization $\sigma_8=0.8$. The density field is sampled with
$480^3$ dark matter particles and an initially equal number of
gas particles, with masses $m_{\rm DM}=4.6\times 10^9 h^{-1} M_\odot$
and $m_{\rm gas}= 6.9\times 10^8 h^{-1} M_\odot$, respectively. The
Plummer-equivalent gravitational softening is 7.5 $h^{-1}$ kpc
comoving at $z>2$, and fixed in physical units at lower redshift.

The simulation was run with {\tt GADGET-2}
\citep*{springel01},
a massively parallel Tree+SPH code with fully adaptive time--stepping,
which uses the energy and entropy conserving SPH
implementation of \citet{springel02}. 
The code includes a photoionizing, time-dependent, uniform UV
background, radiative cooling, star formation, feedback from 
type II supernovae and a phenomenological recipe for
galactic winds. Within each gas particle of sufficiently high density, the gas is a two-phase
fluid, with unresolved cold clouds, embedded at pressure equilibrium
in an ambient hot medium, providing a 
sub-grid model for the multiphase nature of the interstellar
medium \citep{springel03}.

The volume of our simulation is larger than $140^3 h^{-3}$ Mpc$^3$, as required to predict the 
evolution of cluster peculiar velocities correctly \citep{sheth01}; this volume also yields
a cluster sample large enough for statistical purposes. Moreover our simulation
represents a substantial improvement compared to recent cosmological hydrodynamical simulations
used for the study of the SZ effect: they typically have a volume $100^3 h^{-3}$ Mpc$^3$ and a
particle masses four to six times larger than in our simulation (\citealt{white02}; \citealt{dasilva03}).
Finally, the good mass and spatial resolution and the treatment of a two-phase fluid for
star formation and feedback provide a reasonable modelling of the gas physics.

We identify clusters in the simulation box with a two-step procedure:
a friends-of-friends algorithm applied to the dark matter particles alone
provides a list of halos whose centres are used as input to the spherical
overdensity algorithm which outputs the final list
of clusters \citep{borgani03}. Centered on the most bound particle of each
cluster, the sphere with virial overdensity $\Delta_c(z)$, with
respect to the critical density \citep*{eke96}, defines the virial
radius $R_{\rm vir}$.  For this cosmological model $\Delta_c=101$, at
$z=0$.  At redshift $z=0$ the simulation box contains 117 clusters
with mass $M(<R_{\rm vir})\ge 10^{14} h^{-1} M_\odot$. This cluster
set is our sample.

\begin{figure}
\caption{Maps along three orthogonal directions of the thermal SZ
effect for a simulated cluster with mass $M(<R_{\rm vir})=6.95\times
10^{14} h^{-1} M_\odot$.  Each pixel has size $61.6 h^{-1}$ kpc on a
side and the field of view is $7.9 h^{-1}$ Mpc on a side.}
\label{fig:tSZ_map}
\end{figure}  

\begin{figure}
\caption{Same as Figure \ref{fig:tSZ_map} for the kinetic SZ effect.}
\label{fig:kSZ_map}
\end{figure} 

\subsection{Simulated maps}
 
Around each cluster we extract 
a spherical region extending out to $6\,R_{\rm vir}$.
We then create maps of the relevant quantities along three orthogonal 
directions to investigate projection effects. Each map is a regular $N_p\times N_p$ grid. 
In the Tree+SPH code, each
gas particle has a smoothing length $h_i$ and the thermodynamical
quantities it carries are distributed within the sphere of radius
$h_i$ according to the compact kernel
\begin{displaymath}
W(x)={8\over \pi h_i^3}\left\{\begin{array}{ll}
1 - 6x^2 + 6x^3  &  0\le x \le {1\over2} \\
 2(1-x)^3       &         {1\over 2}<x\le 1 \\
  0            &       x\ge 1\; , \end{array} \right.
\end{displaymath}
where $x=r/h_i$ and $r$ is the distance from the particle position.
We therefore distribute the quantity of each particle on the grid
points within the circle of radius $h_i$ centered on the particle.
Specifically, we compute a generic quantity $q_{jk}$ on the grid point
$\{j,k\}$ as $q_{jk} d_p^2= \int q(r){\rm d}l d_p^2 =\sum q_i (m_i/\rho_i)
w_i$ where $d_p^2$ is the pixel area, the sum runs over all the
particles, and $w_i \propto\int W(x){\rm d}l$ is the weight proportional to
the fraction of the particle proper volume $m_i/\rho_i$ which
contributes to the grid point $\{j,k\}$. For each particle, the
weights $w_k$ are normalized to satisfy the relation $\sum w_k=1$
where the sum is now over the grid points within the particle
circle. When $h_i$ is so small that the circle contains no grid point,
the particle quantity is fully assigned to the closest grid point.
Results in this paper are shown for a number of pixels $N_p=128$,
corresponding to a comoving length resolution $\approx 43 h^{-1}$ kpc,
on average.  Figures \ref{fig:tSZ_map} and \ref{fig:kSZ_map} show an
example of the thermal and the kinetic SZ maps of a cluster in our
simulation.

\section{Scaling relations}\label{sec:scaling}

Galaxy clusters are self-similar in their dark matter component, but
not necessarily in their baryonic component. Non-gravitational
processes, namely gas heating and cooling, vary the entropy of the ICM
and can substantially affect the distribution of the gas
(\citealt{borgani03}; \citealt{voit04}). Measuring scaling relations among gas properties
is an important step to understand the role of the different processes
governing the ICM physics. The list of the relevant processes that we
included in our simulation is currently one of the most complete.
Therefore, we can attempt a comparison of our simulated gas properties
with the observations of the SZ effect.

Here, we will investigate the relations $L_X-y_0$ and $T-y_0$ between 
the X-ray luminosity $L_X$ or the ICM temperature $T$ and the central Comptonization parameter 
$y_0$; we will also compute the relation $T-\Delta S$, where $\Delta S$ is the SZ surface brightness 
integrated over a given solid angle. 

The self-similar model of cluster formation predicts power-law scaling relations
between these quantities (e.g. \citealt{evrard91}; \citealt{kaiser91}; 
\citealt{eke98}; \citealt{dasilva03}). 
The cluster virial mass scales as
$M_{\rm vir} \propto \rho_c(z) \Delta_c(z) R^3 $, where $R$ is the cluster 
radius in physical units and $\rho_c(z)$ is the critical
density of the universe; $\rho_c(z)$ scales with redshift $z$ as $\rho_c(z)\propto E^2(z) 
= \Omega_0(1+z)^3 + \Omega_\Lambda$, in a flat universe.
If we assume that clusters have a constant
gas mass fraction and $\Delta_c(z)$ is constant with 
redshift,\footnote{In a flat universe, when $\Omega_0=0.1$, $\Delta_c(z)$ increases by a factor $\sim 2.5$
between, e.g., $z=0$ and $z=5$, and this factor decreases with increasing $\Omega_0$. On the
other hand, $E^2(z)$ increases by 
a factor $ \sim 22$ between the same redshifts, and this factor increases with
increasing $\Omega_0$.} 
the cluster size $R$ scales with $z$ and the ICM mass $M$ 
as $R\propto M^{1/3} E^{-2/3}(z) $. Therefore 
the ICM mass scales with its temperature $T$ as $M E(z)\propto T^{3/2}$;
for the CMB flux variation due to the thermal SZ effect we have $\Delta S \propto 
\int y(\btheta) {\rm d}\Omega \propto d_A^{-2} \int T n_e {\rm d}^3r \propto d_A^{-2} T^{5/2} E^{-1}(z)$,
where $d_A$ is the angular diameter distance to the cluster, and ${\rm d}^3r=d_A^2 {\rm d}\Omega {\rm d}l$. 
We can also write $\Delta S$ in a different way to get its explicit dependence on $y_0$: 
$\Delta S \propto y_0 d_A^{-2} \int {\rm d}A \propto y_0 d_A^{-2} M^{2/3} E^{-4/3}(z)
\propto y_0 d_A^{-2} T E^{-2}(z)$, where ${\rm d}A=d_A^2{\rm d}\Omega$ is the infinitesimal
projected area covered by the cluster. We find $y_0\propto T^{3/2}E(z)$. 
Finally, the X-ray luminosity $L_X\propto \int n_e n_H \Lambda(T) {\rm d}V \propto T^2 E(z)$,
if the cooling function $\Lambda(T)\propto T^{1/2}$, and we find $y_0\propto L_X^{3/4} E^{1/4}(z)$.
In summary, we expect
\begin{equation}
y_0\propto L_X^{3/4} E^{1/4}(z)\; ,
\end{equation}
\begin{equation}
y_0\propto T^{3/2} E(z)\; ,
\end{equation}
\begin{equation}
\Delta S d_A^2 E(z)\propto T^{5/2}\; ,
\end{equation}
if the self-similar model holds.

\begin{figure}
\includegraphics[scale=0.38,angle=90]{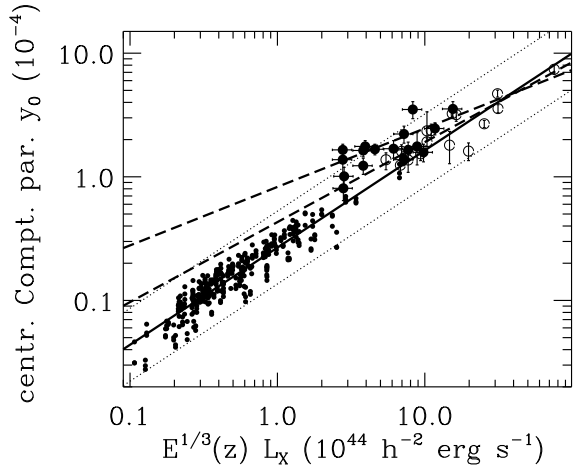}
\caption{Central peak of the
Comptonization parameter map vs. the X-ray bolometric luminosity.  The dots are the simulated
clusters, the open and solid circles with error bars the cluster sample of \citet{mccarthy03b}
and \citet{cooray99}, respectively.  
The solid line is the best fit to the simulated clusters and the
two dotted lines show the $\pm 3\sigma$ range from the best fit.
The two dashed lines are the best fits to the observed samples. The steeper
dashed line is the fit to the sample of \citet{mccarthy03b}.}
\label{fig:LX-vs-y}
\end{figure}

Figure \ref{fig:LX-vs-y} shows the correlation between 
the central peak $y_0$ of the Comptonization parameter map and the X-ray
bolometric luminosity $L_X$. 
Since the simulation only includes an approximate treatment of
metal production and the cooling function used in the code is computed for zero
metallicity, we
follow \citet{borgani03} and consider an optically thin gas
of primordial cosmic abundance $X=0.76$ and $Y=0.24$ to compute the
ion number density ${n_{\rm H}}_i + {n_{\rm He}}_i$.  The X-ray
luminosity of a simulated cluster is thus
\begin{equation}
L_X = \sum {n_e}_i ({n_{\rm H}}_i + {n_{\rm He}}_i) \Lambda(T_i) dV_i
\end{equation}
with the appropriate cooling function $\Lambda(T)$ \citep{suth93};
$dV_i=m_i/\rho_i$ is the volume occupied by 
the gas particle of mass $m_i$, density $\rho_i$ and electron number 
density ${n_e}_i$, $T_i$ is its temperature and the
sum runs over the gas particles, within $R_{\rm vir}$, with
$T_i>3\times 10^4$ K and $\rho_i$ smaller than 500 times
the mean baryon density (\citealt{croft01}; \citealt{borgani03}). 
The sum therefore does not include dense
star-forming particles, which are regulated by the multiphase
model and may produce, when included, a spurious, highly uncertain contribution
to the X-ray emission.

\begin{figure}
\includegraphics[scale=0.38,angle=90]{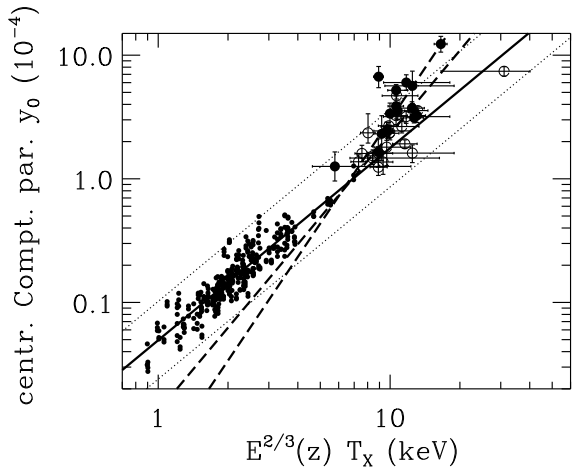}
\caption{Central 
peak of the Comptonization parameter map vs. the X-ray emission weighted temperature.
The steeper dashed line is the best fit to the sample of \citet{benson04} (solid circles with error bars). 
Other symbols and lines are as in Figure \ref{fig:LX-vs-y}.
The temperatures of the real clusters are corrected for the presence of a cooling flow.}
\label{fig:TX-vs-y}
\end{figure}   

\begin{figure}
\includegraphics[scale=0.38,angle=90]{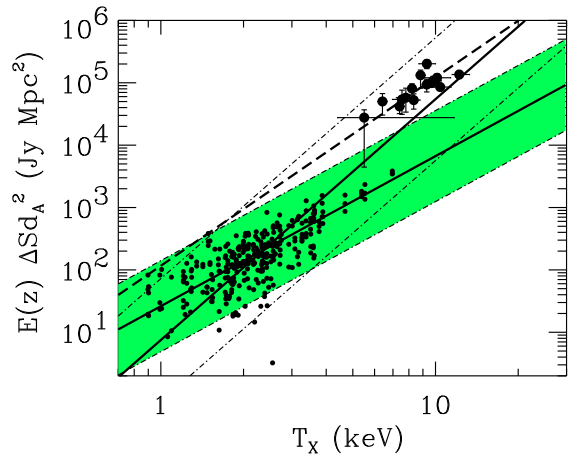}
\caption{SZ flux decrement at 145 GHz, integrated within $r_{2500}$, vs. 
the X-ray emission weighted temperature. 
Symbols and the dashed line are as in Figure \ref{fig:TX-vs-y}.
The solid lines are the fits to the simulated sample using the least absolute
deviation method. Dot-dashed lines show the range $\pm 3D$  
where $D$ is the mean absolute deviation, from the fitting relation, of each simulated cluster 
in the $\Delta S$-direction.
The shallower relation, which corresponds to the shaded area, assumes $T_X$  
as the independent variable, whereas
the steeper relation assumes $E(z) \Delta S d_A^2$ 
as the independent variable. The $\pm 3D$ range of the shallower relation has been shaded for clarity.
The temperatures of the real clusters are not corrected for the presence of a cooling flow.}
\label{fig:y-flux-vs-TX}
\end{figure}
            
The best fit to the simulated cluster sample (the dots in Figure \ref{fig:LX-vs-y}) is
\begin{eqnarray}
\log\left(y_0\over 10^{-4}\right) & = & 
(-0.57\pm 0.01) + \cr 
 & + &(0.79\pm 0.02) \log\left[L_X\over 10^{44} h^{-2} {\rm erg\ s}^{-1}\right] \; ,
\end{eqnarray}
in agreement with the self-similar model.
We do not attempt to simulate observational error bars on the cluster data extracted
from the simulation, because such a detailed analysis is beyond the scope of this paper.  
Therefore, in the absence of errors on the simulated data,
the uncertainties on the fitting parameters we compute are only indicative 
(e.g. \citealt{press92}, p. 661).
An alternative estimate of the uncertainty on this scaling relation
is provided by the relative variance on a logarithmic scale 
$\sigma^2= \langle[\log(y_0)-\log(y_0^{\rm fit})]^2\rangle$. We find $\sigma=0.10$,
comparable to the typical uncertainties of observed $y_0$'s. The dotted lines 
in Figure \ref{fig:LX-vs-y} show the $\pm 3\sigma$ amplitude.

An analogous fit to 17 observed clusters (open circles in Figure \ref{fig:LX-vs-y}) drawn 
from the sample compiled by \citet{mccarthy03b}, 
mostly based on the sample of \citet{reese02}, yields the relation:
$\log(y_0/10^{-4}) = (-0.37\pm 0.05) + (0.65\pm 0.04) \log[E^{1/3}(z)L_X/ 10^{44}h^{-2} {\rm erg\ s}^{-1}]$,
in reasonable agreement with the self-similar model and our simulated sample.
In computing $E(z)$, we assume $\Omega_0=0.3$ and $\Omega_\Lambda=0.7$.


In passing, we note that the apparent
bimodal distribution between simulated and observed data is a
consequence of the observational bias of pointing at the very
massive clusters: the volume of our simulation box is not large enough
to include that many massive clusters.

Despite their small statistical errors, the fit parameters on the real sample should be treated
with caution.
In fact, the sample of \citet{cooray99} (solid circles) 
yields a shallower relation 
$\log(y_0/10^{-4}) = (-0.08\pm 0.05) + (0.47\pm 0.07) \log[E^{1/3}(z) 
L_X/ 10^{44}h^{-2} {\rm erg\ s}^{-1}]$.\footnote{Wherever appropriate, in computing the best fit
we consider the uncertainties on both axes (e.g. \citealt{press92}, p. 666).}
The discrepancy between the two observed samples might be statistically meaningless, however, because 
the probability $q$ of getting a larger $\chi^2$
with a sample drawn from a cluster population which obeys either fit to the observed samples 
is $\la 4\times 10^{-8}$ (Table \ref{tab:fit-params}).
The small uncertainties claimed for the observed values 
explain the large $\chi^2$'s, which on turn originate the small significance
levels $q$. The discovery of possible systematic errors can of course bring the observed samples in 
better agreement with each other and with a single scaling relation.
Indeed the \citet{cooray99} sample has 10 out of 14 clusters in common with \citet{mccarthy03b} sample,
but this latter sample have more recent and different measures of $y_0$ and, in some cases, of $L_X$: these differences 
might reflect both random and systematic errors.

An intrinsic scatter in the $y_0-L_X$ scaling relation is in any case expected, due to the simplifying assumptions used to derive
$y_0$, namely the isothermality of the ICM, the $\beta-$model and the cluster sphericity.  It is remarkable that the scatters shown
by our simulated clusters and by the observed samples are comparable.

 \begin{table}
 \caption{Fit parameters to the scaling relations $\log(f_2Y)=a+b\log(f_1X)$.}
 \label{tab:fit-params}
 \begin{tabular}{@{}llcccc}
  $X-Y$ & & $a$ & $b$ & $\chi^2/\nu$ & $q$ \\
 \\
  $L_X-y_0$  & (1) & $-0.08\pm 0.05$ & $0.47\pm 0.07$  & 4.09 & 2.5e-8 \\
  $f_1^3=E$	    & (2) & $-0.37\pm 0.05$  & $0.65\pm 0.04$ & 4.31 & 3.9e-8 \\ 
  $f_2=1$	    & (4) & $-0.57\pm 0.01$  & $0.79\pm 0.02$ & {} & {} \\ 
 \\
  $T_X-y_0$ & (1) & $-1.60\pm 0.31$  & $1.87\pm 0.31$ & 1.78 & 2.5e-2 \\ 
  $f_1^3=E^2$	    & (2) & $-1.88\pm 0.39$ & $2.24\pm 0.39$ & 1.06 & 0.39 \\
  $f_2=1$	    & (3) & $-2.31\pm 0.54$ & $2.79\pm 0.51$ & 3.03 & 1.7e-4 \\
  {} 	 	    & (3a) & $-2.60\pm 0.59$ & $3.27\pm 0.60$ & 6.94 & 1.3e-13 \\
  {}	    & (4) & $-1.31\pm 0.01$ & $1.55\pm 0.03$ & {} & {} \\
 \\
  $T_X-\Delta S$ & (3) & $2.73\pm 0.38$ & $2.26\pm 0.38$ & 0.70 & 0.76 \\ 
  $f_1=1$        & (3a) & $2.07\pm 0.42$ & $3.02\pm 0.44$ & 2.30 & 4.9e-3 \\
  $f_2=d_A^2E$	    & (4) & $1.42\pm 0.04$ & $2.41\pm 0.11$ & {} & {} \\
  {}	    & (4a) & $0.88\pm 0.12$ & $3.86\pm 0.19$ & {} & {} \\

 \end{tabular}

 \medskip
 (1) \citet{cooray99}; (2) \citet{mccarthy03b}; 
(3)/(3a) \citet{benson04} with temperatures corrected/not corrected
for the presence of a cooling flow; (4) simulated clusters; in the $T_X-\Delta S$ relation the fits
to the simulated sample are computed with the method of the least mean absolute deviation; 
(4a) is the fit to the simulated sample assuming $E(z)\Delta S d_A^2$, rather than $T_X$, 
as the independent variable. Units: $L_X$: $ 10^{44} h^{-2} {\rm erg\ s}^{-1}$; 
$T_X$: keV; $y_0$: $10^{-4}$; $\Delta S d_A^2 $: Jy Mpc$^2$.
$\nu$ is the number of degrees of freedom; $q$ is the significance level of 
$\chi^2/\nu$.    
\end{table}

Discrepancies among observed samples can also be seen in the relation
between $y_0$ and the X-ray temperature $T_X$ (Figure \ref{fig:TX-vs-y}).
For the simulated clusters we plot the emission--weighted
temperature averaged over the gas particles within $R_{\rm vir}$
\begin{equation}
T_X = {\sum {n_e}_i ({n_{\rm H}}_i + {n_{\rm He}}_i) \Lambda(T_i) T_i dV_i \over
\sum {n_e}_i ({n_{\rm H}}_i + {n_{\rm He}}_i) \Lambda(T_i) dV_i}\; . 
\label{eq:sim-TX}
\end{equation}

The simulated clusters yield the relation
\begin{equation}
\log\left(y_0\over 10^{-4}\right) = (-1.31\pm 0.01) + (1.55\pm 0.03) \log\left[T_X\over {\rm keV}\right],
\end{equation}
as expected in the self-similar model.
For this relation, the standard deviation on a logarithmic scale is 
$\sigma= \langle[\log(y_0)-\log(y_0^{\rm fit})]^2\rangle^{1/2}=0.10$ 
(the dotted lines in Figure \ref{fig:TX-vs-y}
show the $\pm 3\sigma$ range). 

The sample of \citet{cooray99} yields a relation close to the self-similar prediction:
$\log(y_0/ 10^{-4}) = (-1.60\pm 0.31) + (1.87\pm 0.31) \log[E^{2/3}(z)T_X/ {\rm keV}]$, whereas the  
sample of \citet{mccarthy03b} and the more recent sample of \citet{benson04}, which has 12 out of 15 clusters in common
with the \citet{mccarthy03b} sample, yield considerably steeper relations (Table \ref{tab:fit-params}).

Some observed clusters have a central cooling flow.
The temperatures of real clusters shown in Figure \ref{fig:TX-vs-y} are corrected 
for this effect. Neglecting this correction
generally decreases the estimate of $T_X$ and can substantially change the
fit parameters (Table \ref{tab:fit-params}). 
From the statistical point of view, however, we can again draw very little conclusions
from most of these $T_X-y_0$ fits: only the \citet{mccarthy03b} sample yields a large
 significance level $q=0.39$.

As we mentioned earlier, 
the discrepancies that we find among the different observed samples
may be due to systematic errors affecting the data analysis.
Indeed, \citet{benson04} have recently shown how the central $y_0$'s calculated from the 
measurements made with the SuZIE II receiver can be up to 60 per cent
higher than the values derived from the BIMA and OVRO interferometers. 
Therefore, \citet{benson04} prefer to use the SZ flux decrement,
rather than the central $y_0$, as an indicator of the magnitude of the SZ effect. 

Figure \ref{fig:y-flux-vs-TX}
shows the relation $T_X-\Delta S$, where $\Delta S$ is the SZ flux within the circle 
of radius $r_{2500}$ centered on the cluster: the
average mass density within the sphere of radius $r_{2500}$ is 
2500 times the critical density. According to \citet{benson04}, this radius 
is a suitable choice for the instrumental properties of SuZIE II; 
moreover, $r_{2500}$ is used in X-ray analyses of clusters \citep*{evrard96}.
The simulated sample shows a rather sparse distribution in the 
$T_X-\Delta S$ plane. Therefore we use the least
absolute mean deviation method, which, in this case, is more appropriate than the least square fit method,
to compute the fitting relation.  We find 
\begin{equation}
\log\left[\Delta S d_A^2 \over {\rm Jy\; Mpc}^2\right] = (1.42\pm 0.04) + (2.41\pm 0.11) \log\left(T_X\over {\rm keV}\right),
\end{equation}
as expected in the self-similar model. The observations agree with the self-similar slope only when
we use the cooling flow corrected temperatures (sample 3 in Table \ref{tab:fit-params}). In this case, however,
the normalization of the observed sample is substantially larger than the model. 
Using the temperatures
not corrected for the presence of the cooling flow, as shown in Figure \ref{fig:y-flux-vs-TX} 
(sample 3a), yields a lower normalization but 
a substantially different slope.
We do not have, however, error bars on our simulated data. Therefore, 
the fit is sensitive to the choice of the independent variable. Assuming 
$E(z)\Delta S d_A^2$ as the independent variable in the simulated sample (sample 4a) does indeed yield a substantially 
steeper slope; moreover,  
the sample variance is large enough to compensate the factor of two difference in the normalization. 

In conclusion, our simulated cluster sample seems to provide a reasonable description
of the currently observed scaling relations between SZ and X-ray properties.
To derive stronger constraints on our simulations of the ICM, however, we should perform more realistic
mock observations of SZ clusters, although the disagreement, that we find among the observed samples 
of SZ clusters, casts some doubts on the robustness of the observed relations available to date.

\section{The basic assumptions on the velocity measurements}\label{sec:assumptions}

The measure of the line-of-sight peculiar velocity of clusters with
the SZ effects relies on two fundamental assumptions about the
physical properties of the ICM: (1) bulk flows of gas clouds within the ICM
are negligible and gas and dark matter have the same
mean peculiar velocity, and (2) the ICM temperature estimated from the X-ray
spectrum is a reliable estimate of the temperature weighted by the
electron number density.  We now check on these assumptions on turn.

\subsection{Velocity bias and internal bulk flows}\label{sec:vbias}

We might expect that turbulence induced by mergers, gas cooling,
galactic winds and stellar feedback can provide substantial
contribution to the bulk flow of gas clouds within the ICM (e.g. \citealt{inogamov03};
\citealt*{sunyaev03}; \citealt{schuecker04}). In
clusters with low peculiar velocity, these effects can also result in
a gas mean velocity quite different from the mean velocity of the dark
matter halo.  Our simulation, which includes these physical processes,
should provide a sample of realistic clusters for the investigation of
these effects.

In terms of velocity magnitudes, for each simulated cluster, the three
components of the mean velocity of the gas particles within $R_{\rm
vir}$, $v_{\rm gas}$, never deviate substantially from the
corresponding components of the mean velocity of the dark matter
particles, $v_{\rm DM}$: we find a mean absolute deviation, a more robust
estimator of the width of a distribution than the standard deviation,
$\langle\vert v_{\rm gas}- v_{\rm DM}\vert \rangle=18\pm 15$ km
s$^{-1}$; the maximum value in our sample is $\vert v_{\rm gas}-
v_{\rm DM}\vert_{\rm max}=88$ km s$^{-1}$.

However, in terms of relative velocities, the bias can be substantial.
Setting a threshold at $v_{\rm DM}^t=100$ km s$^{-1}$, we find a mean
deviation $\langle\delta\rangle=\langle\vert (v_{\rm gas}- v_{\rm
DM})/ v_{\rm DM}\vert\rangle= 0.9\pm 2.2$ for the clusters with
$v_{\rm DM}<v_{\rm DM}^t$, an order of magnitude larger and with a substantially
larger deviation than
$\langle\delta\rangle=0.09\pm 0.08$ for the clusters with $v_{\rm
DM}>v_{\rm DM}^t$. We conclude that we should not expect any relevant
systematics from velocity bias when we consider clusters with
line-of-sight peculiar velocities larger than $\sim 100$ km s$^{-1}$.

\begin{figure}
\includegraphics[scale=0.38,angle=90]{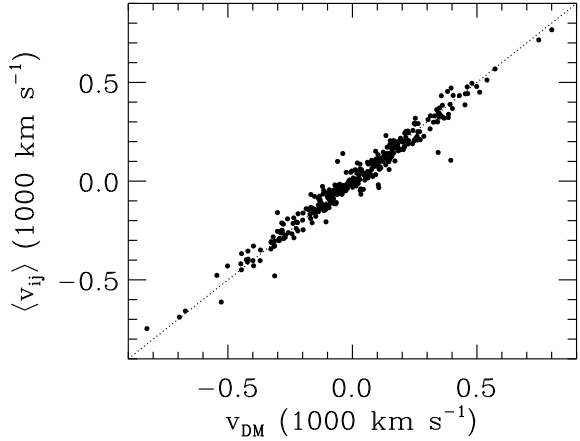}
\caption{Average $\langle v_{ij}\rangle$ of the  velocities at the pixel
sites vs. the peculiar velocity of the cluster $v_{\rm DM}$.
Only pixels within $R_{\rm vir}$ of the cluster centre are included in 
the calculation of the means.
The dotted line is the relation $\langle v_{ij}\rangle=v_{\rm DM}$.}
\label{fig:vmap-vdm}
\end{figure}

We also need to quantify how much the bulk flow of gas clouds within
the ICM can affect the peculiar velocity estimate. We consider the
velocity maps $v_{ij}=c\beta_{ij}$, where $\beta_{ij}$ is a generic element of the array 
corresponding to the map 
computed according to equation (\ref{eq:bel}); therefore, the velocity $v_{ij}$ is 
a velocity weighted by the electron number density.
We first compute the mean $\langle
v_{ij}\rangle$ of the $v_{ij}$ values on the pixels within $R_{\rm
vir}$ of the cluster centre.  This mean agrees with the true $v_{\rm
DM}$ (Figure \ref{fig:vmap-vdm}): the mean absolute deviation is
$\langle\vert \langle v_{ij}\rangle -v_{\rm DM}\vert\rangle= 31\pm 32$
km s$^{-1}$.

\begin{figure}
\includegraphics[scale=0.38,angle=90]{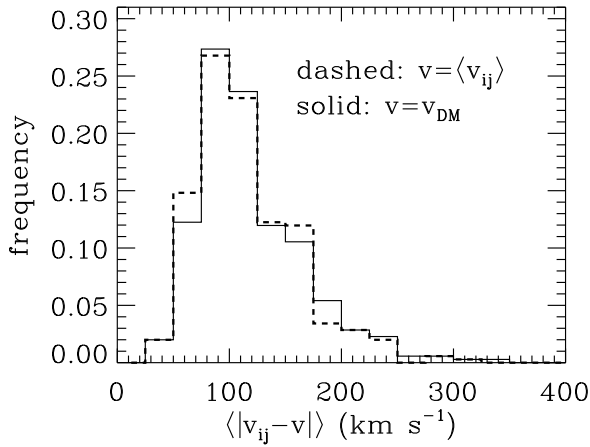}
\caption{Distribution of the mean absolute deviations between the
velocities at the pixel sites and the peculiar velocity of the cluster
$v=v_{\rm DM}$ (solid histogram) or the mean velocity of the velocity
map $v=\langle v_{ij}\rangle$ (dashed histogram).  Only pixels within
$R_{\rm vir}$ of the cluster centre are included in the calculation of the means.
}
\label{fig:vmapscatter}
\end{figure}

For each map, we then compute the mean absolute deviation
$\langle\vert v_{ij}-v\vert\rangle$ of the velocity values at the
pixel sites from $v=v_{\rm DM}$ or $v=\langle v_{ij} \rangle$.  Figure
\ref{fig:vmapscatter} shows the distribution of the $\langle\vert
v_{ij}-v\vert\rangle$'s for all the maps. The average values are
$\langle\vert v_{ij}-v_{DM}\vert\rangle = 119\pm 48 $ km s$^{-1}$ and
$\langle\vert v_{ij}-\langle v_{ij}\rangle\vert\rangle = 115\pm 45 $
km s$^{-1}$.  We do not expect the two distributions to be different
because $\langle\vert \langle v_{ij}\rangle -v_{\rm DM}\vert\rangle$
is small.  The two samples, with $v_{\rm DM}$ smaller or larger than
the threshold $100$ km s$^{-1}$, have a $\langle\vert
v_{ij}-v\vert\rangle$ distribution similar to the complete sample,
both when $v=v_{\rm DM}$ and $v=\langle v_{ij} \rangle$.  This last result
indicates that the velocity bias found above for the slow clusters
($\langle\delta\rangle= 0.9\pm 2.2$) is totally due to the bulk flow
of gas clouds within the ICM, and not to different mean velocities 
of gas and dark matter particles. This conclusion also explains
why the scatter of $\langle\delta\rangle$ for 
the slow clusters ($2.2$) is much larger than the scatter for
the fast clusters ($0.08$): bulk flows have roughly the same
magnitude, independently of the cluster speed. 

According to Figure \ref{fig:vmapscatter}, internal bulk flows
produce a deviation, from $v_{\rm DM}$, smaller than 200 km s$^{-1}$ 
in 93 per cent of the cases.
We conclude that the velocity bias and internal bulk flows
set a conservative lower limit of $\sim 200$ km s$^{-1}$ to 
the uncertainty on peculiar velocity measures
with the SZ effects (\citealt{nagai03}; \citealt{hold03}), 
well below the uncertainties of the currently most accurate observations
(\citealt{laroque03}; \citealt{benson03}; \citealt{kitayama04}),
but similar to those expected in future measurements at microkelvin sensitivity
and arcminute angular resolution \citep{knox04}.


\begin{figure}
\includegraphics[scale=0.38,angle=90]{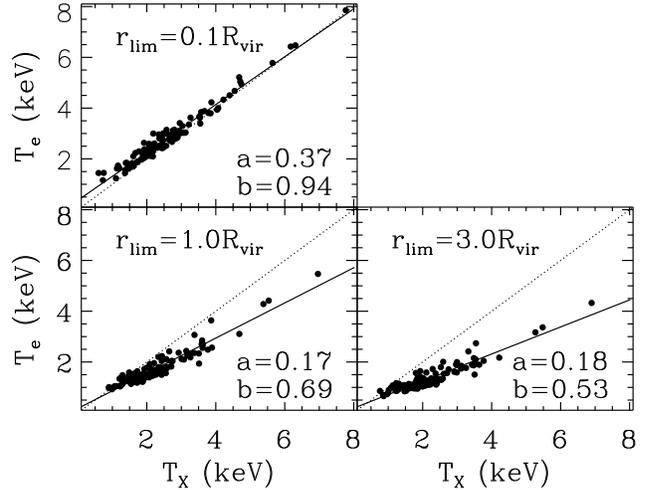}
\caption{Electron density weighted temperature $T_e$ vs. X-ray
emission--weighted temperature $T_X$.  Only gas particles within
$r_{\rm lim}$ of the cluster centre are considered. 
Each panel shows the $T_X-T_e$ relation for a different value of $r_{\rm lim}$, as indicated. 
The dotted line is the equation $T_e=T_X$; the solid line is the best fit
$T_e=a+bT_X$. The parameter $a$ is in keV.}
\label{fig:Te-TX}
\end{figure}    

\begin{figure}
\includegraphics[scale=0.38,angle=90]{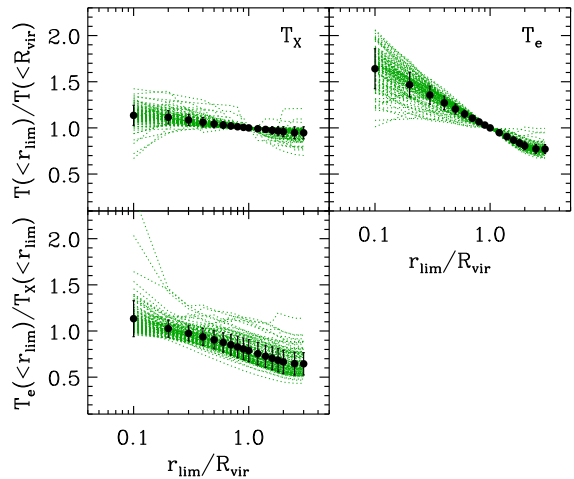}
\caption{Temperature profiles of the simulated clusters.
The upper left (right) panel shows the temperature weighted by the X-ray emission (electron number density).
The bottom panel shows the profile of the ratio of the two temperatures.
Dotted lines are the individual profiles. Dots with error bars show the mean
profiles with 1-$\sigma$ standard deviations.}
\label{fig:T-prof}
\end{figure}

\subsection{Electronic and X-ray temperatures}\label{sec:Te-TX}

For all X-ray emitting particles, we can compute an X-ray emission 
weighted temperature (equation \ref{eq:sim-TX}) and an electron density weighted temperature 
\begin{equation}
T_e = {\sum {n_e}_i T_i dV_i\over \sum {n_e}_i dV_i }\; .
\label{eq:Telpart}
\end{equation}
In both equations (\ref{eq:sim-TX}) and (\ref{eq:Telpart}), the 
sums run only over the particles within a maximum radius $r_{\rm lim}$.

Note that $T_e$ equals a mass-weighted temperature $T_M=\sum m_i T_i / 
\sum m_i $
when the gas is fully ionized, namely ${n_e}_i\propto \rho_i$, 
which is usually the case for temperatures larger than 0.1 keV.

Figure \ref{fig:Te-TX} shows $T_e$ vs. $T_X$ for different values of
$r_{\rm lim}$. The two temperatures are comparable only at
the smallest $r_{\rm lim}$, when $T_e$ is roughly 10 per cent larger than 
$T_X$ \citep{mathiesen01}. At larger $r_{\rm lim}$'s the disagreement 
increases. In fact, the ICM temperature  
profile drops at large radii both in real clusters (e.g. \citealt{degrandi02}; \citealt{kaastra03})
and in simulations (e.g. \citealt{lin03}; \citealt{borgani03}; \citealt*{rasia04b}). 

Figure \ref{fig:T-prof} shows the profiles of our simulated clusters.
$T_X$ is largely insensitive to $r_{\rm lim}$ because
it is proportional to the square of the gas density and 
is therefore dominated by the emission from the central region.
Instead, $T_e$ steeply decreases with increasing $r_{\rm lim}$ 
because it is only proportional to the first power of the electronic gas density. 

When $r_{\rm lim}=R_{\rm vir}$, $T_e$ can be
substantially smaller than $T_X$: the mean relation between these two
measures of the ICM temperature is
$T_e/{\rm keV} =0.17+0.69(T_X/{\rm keV})$ (Figure \ref{fig:Te-TX}). 
Therefore, in spatially poorly resolved clusters, 
using $T_X$ rather than $T_e$ can substantially
overestimate the peculiar velocity (equations \ref{eq:vel-res} and
\ref{eq:vel-unres}), as we will see in the next section.

Note that here we compute the bolometric temperature. Figure 3 in
\citet{borgani03} shows that the temperatures estimated by weighting
the emissivity in a finite energy band are larger than those obtained by weighting with
the bolometric emissivity.  Therefore, for
real clusters, the $T_e-T_X$ relation can be shallower and the
overestimate of the peculiar velocity more severe. It remains to quantify this trend
when a realistic observational procedure is applied to simulated clusters \citep{mazzotta04}.

\begin{figure*}
\includegraphics[scale=0.67,angle=90]{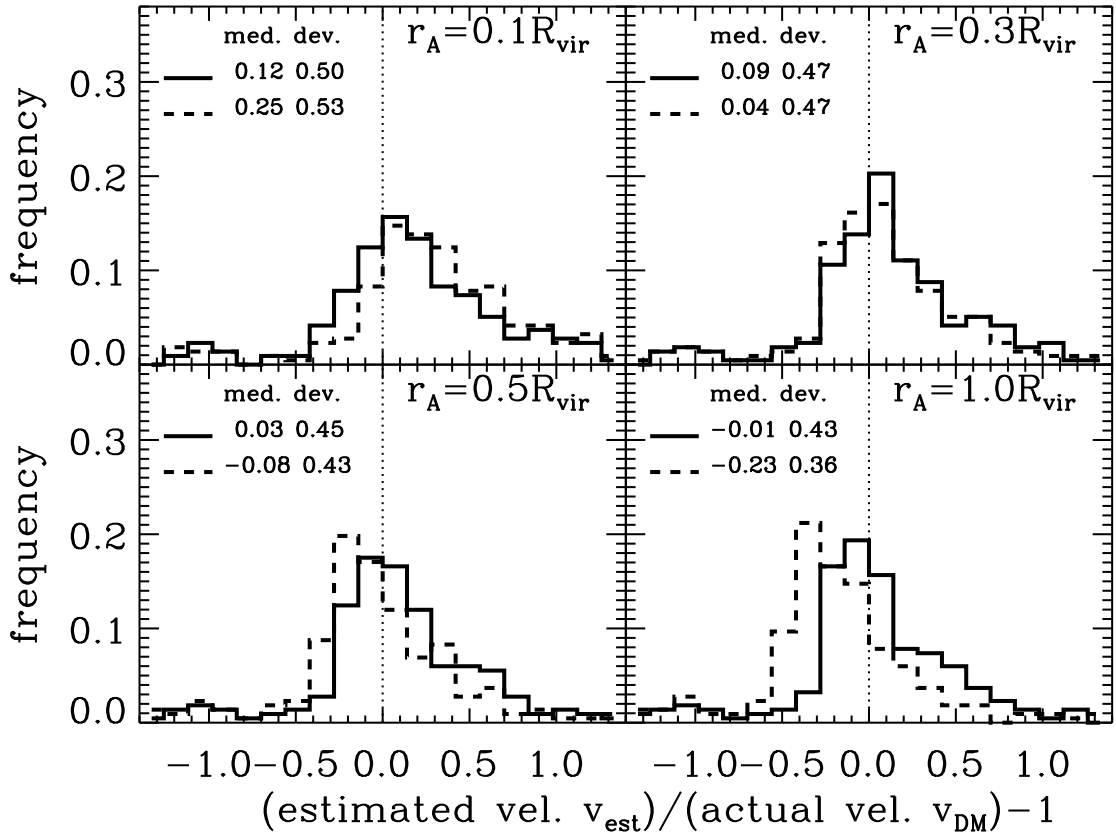}
\caption{Distributions of the deviations of the estimated line-of-sight velocities from
the real velocities of spatially resolved clusters with $v_{DM}>100$ km s$^{-1}$. 
The velocities are estimated with equations (\ref{eq:vel-res}) where
$T_e =\langle T_X\rangle_{r_A}$ (solid histograms),
or $T_e = \langle T_e\rangle_{r_A}$ (dashed histograms) and the $\beta$-model parameters
are derived from the fit to the X-ray surface brightness. 
Each panel shows the distributions for a different value of $r_A$, as indicated.
The other numbers in each panel are the median and the mean absolute deviation
from the median of each distribution.}
\label{fig:vel-Te-TX-dist-beta}
\end{figure*}

\begin{figure*}
\includegraphics[scale=0.67,angle=90]{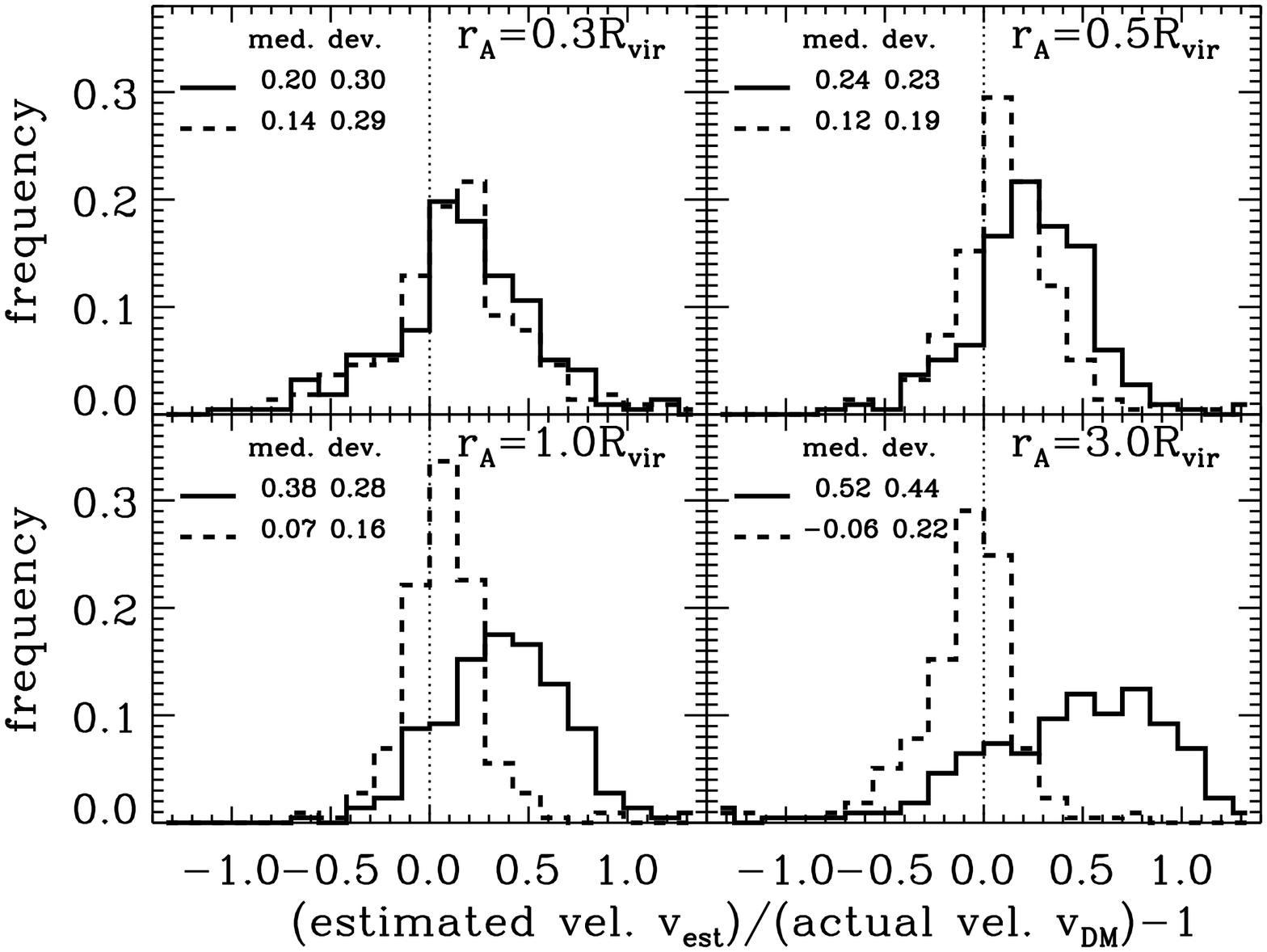}
\caption{Distributions of the deviations of the estimated line-of-sight velocities from
the real velocities of spatially unresolved clusters with $v_{DM}>100$ km s$^{-1}$.
The velocities are estimated with equations (\ref{eq:vel-unres}) where
$\langle T_e\rangle_{\theta_b} =\langle T_X\rangle_{r_A}$ (solid histograms),
or $\langle T_e\rangle_{\theta_b} = \langle T_e\rangle_{r_A}$ (dashed histograms).
Each panel shows the distributions for a different value of $r_A$, as indicated.
The other numbers in each panel are the median and the mean absolute deviation
from the median of each distribution.}
\label{fig:vel-Te-TX-dist-apert}
\end{figure*}

\section{Peculiar velocity measurements}\label{sec:velocity}

Measuring the peculiar velocity of clusters with the SZ effect 
requires the separation of the thermal from the kinetic contribution to the
total SZ flux. Spatially resolved and unresolved clusters 
require two different procedures for this separation. 

In the former, a model for the ICM distribution is derived from the X-ray 
surface brightness to estimate
the central values of the SZ contributions. In the latter, the beam-averaged
SZ contributions are measured without any information on 
the ICM distribution. To date,
only the former case has been applied to observations.

We now consider these two approaches on turn to estimate
the systematics that they introduce on the peculiar velocity measurement. 
Here, we will only consider the physical properties of the ICM and we will not model 
the instrumental or data reduction problems 
which affect real observations.

\subsection{Resolved clusters: fitting profile method}

We perform a rather simple simulation of the observation
of individual clusters.
Namely, we (1) construct an X-ray surface brightness map from the simulation,
(2) derive the gas distribution shape parameters $\beta_c$ and $\theta_c$
from the X-ray surface brightness profile,
(3) compute the SZ fluctuations in the CMB intensity maps $\Delta I(\nu)$ 
at four different frequencies $\nu=30$, $143$, $217$, and $353$ GHz, 
(4) derive the central values $\Delta I_0(\nu)$ of the radial fits
with the shape parameters $\beta_c$ and $\theta_c$
imposed by the X-ray surface brightness profile, (5) perform a fit
to equation (\ref{eq:DeltaI}) by assuming that the four $\Delta I_0(\nu)$'s are independent,
(6) estimate the peculiar velocity (equation \ref{eq:vel-res}) by using the X-ray temperature
$T_X$ (equation \ref{eq:sim-TX}) or the electronic temperature $T_e$ (equation \ref{eq:Telpart}).

In equation (\ref{eq:vel-res}), $T_e$ indicates the central ICM temperature
weighted by the electron density. Observationally, one uses the X-ray
temperature averaged within the resolution beam. 
To quantify the effect of the beam size, we compute the temperature with 
the gas particles within spheres of different radii $r_A$. 

Figure \ref{fig:vel-Te-TX-dist-beta} shows the distribution
of the deviations $(v_{\rm est}-v_{\rm DM})/v_{\rm DM}$
of the estimated line-of-sight peculiar velocity $v_{\rm est}$ from the
actual cluster velocity $v_{\rm DM}$. The solid and dashed histograms
show these deviations when we use $T_X$ or $T_e$, respectively.
We consider the temperatures averaged within four different radii. 

The temperature within the smallest radius $0.1R_{\rm vir}$ is 
the closest to the central temperature that should be used in equation (\ref{eq:vel-res}).
At this radius, both the X-ray and the electronic temperature overestimate 
the peculiar velocity. The overestimate with $T_e$ is more severe, because $T_e$ is 
10 per cent larger than $T_X$ (Figure \ref{fig:T-prof}). The incorrect estimate of the peculiar
velocity derives from the fact that the gas is neither spherically
and smoothly distributed nor isothermal, as assumed when we impose the
$\beta$-model to the X-ray surface brightness. In fact, the
agreement between estimated and real peculiar velocity  does not improve when 
we use the $\beta$-model parameters derived from the SZ surface brightness, rather
than the X-ray surface brightness, despite the fact that the two parameter sets 
do not generally coincide (e.g. \citealt*{yoshikawa98}; \citealt{lin03}). 

The radial profile of the X-ray temperature $T_X$ decreases less rapidly
than the radial profile of $T_e$ (Figure \ref{fig:T-prof}). 
Therefore, the peculiar velocity estimate based on $T_X$ is rather insensitive
to the beam size, whereas the velocity estimated with $T_e$ rapidly decreases
with $r_A$.

The distributions in Figure \ref{fig:vel-Te-TX-dist-beta} only include the clusters with $v_{\rm DM}>100$ km s$^{-1}$.
Including all the clusters leaves the medians unchanged, but substantially extends 
the tails of the distributions, because, in slow clusters, errors of order a few tens of km s$^{-1}$ produce 
large relative deviations due to the bulk flows internal to the ICM.
  
\begin{figure}
\includegraphics[scale=0.38,angle=90]{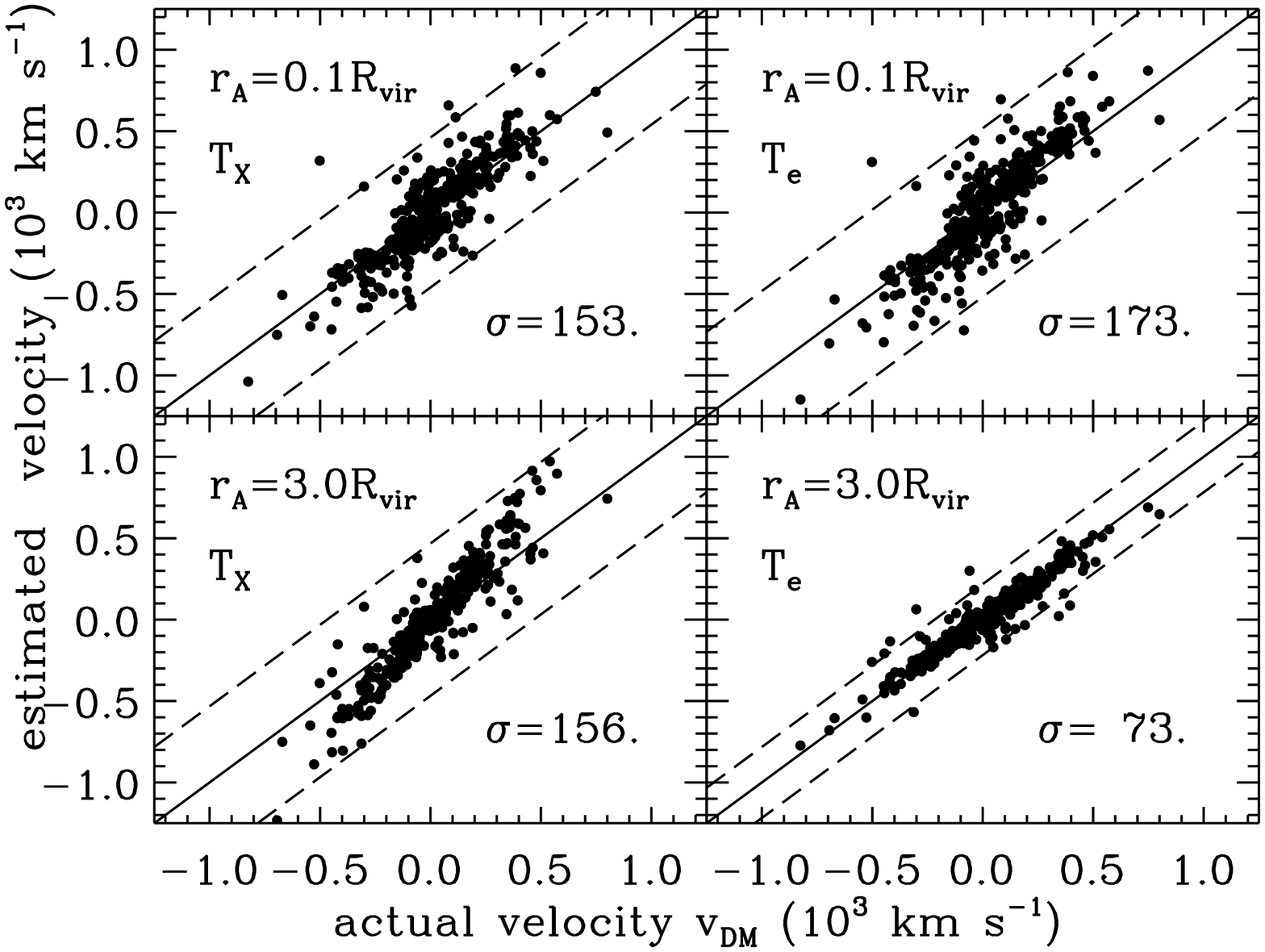}
\caption{Relation between the estimated and the actual velocity.
Upper (lower) panels show spatially resolved (unresolved) clusters.
In left (right) panels the ICM temperature is weighted by the X-ray emission (electron number density)
of the gas particles within $r_A$.  
The solid line is the $v_{\rm est}=v_{\rm DM}$ relation.
The dashed lines indicate the $\pm 3\sigma$ range
from the $v_{\rm est}=v_{\rm DM}$ relation, where $\sigma^2=\langle(v_{\rm DM}-v_{\rm est})^2\rangle$.
The numbers in each panels are the standard deviations $\sigma$ in km s$^{-1}$.}
\label{fig:vel-TX-Te-res-unres}
\end{figure}

In terms of velocity magnitudes, the deviations are comparable, on average, to the systematics due 
to the internal bulk flows. The upper panels in Figures \ref{fig:vel-TX-Te-res-unres} show
the scatter plot between the estimated and the real velocities for all the clusters when 
$r_A=0.1R_{\rm vir}$.
The standard deviations are $\sigma=\langle (v_{\rm est}-v_{\rm DM})^2\rangle^{1/2}=153$ 
km s$^{-1}$ and 173 km s$^{-1}$, when we use the X-ray or the electronic temperatures, respectively. 

If our simulated clusters mirror a realistic sample, we expect that,
for spatially resolved clusters with peculiar velocity larger than $\sim 100$ km s$^{-1}$, 
the usual assumptions on the ICM physical properties should cause
on average a $\sim 10-20$ per cent systematic overestimate of the velocity. 
For these clusters, the width of the distribution in the upper-left panel of Figure \ref{fig:vel-Te-TX-dist-beta}
suggests $\sim 50$ per cent as a typical uncertainty on the velocity estimate.

\subsection{Unresolved clusters: aperture method}

When the radio/microwave observations do not spatially resolve the cluster, 
we need to use the equations described in Sect. \ref{sec:unresolved}.
Here, we compute the $y(\btheta)$ and $b(\btheta)$ maps, and 
compute $\Delta S(x)$ by integrating $\Delta I(x)$ over a beam of physical size 
$2r_A(\theta_b)=d_A\theta_b$.
We then extract $Y(\theta_b)$ and $ B(\theta_b)$ 
from the spectral fit. We emphasize that in this case there is no need of 
any assumption about the ICM spatial distribution.  

Figure \ref{fig:vel-Te-TX-dist-apert} shows the distributions
of the deviations of the estimated line-of-sight velocities from the real velocities
for four different aperture radii $r_A$. Our maps have $128\times 128$ pixels 
and are $12 R_{\rm vir}$ on a side. Therefore, we choose $r_A\ge 0.3R_{\rm vir}$
to avoid the shot-noise due to the small number of pixels in smaller apertures. 

When we use the X-ray temperature averaged over the appropriate aperture 
in equation (\ref{eq:vel-unres}), the larger is the aperture, the larger
is the line-of-sight peculiar velocity overestimate (solid histograms). 
In fact, the ratio of the two integrated SZ components $B(\theta_b)/Y(\theta_b)$,
in units of the central ratio $b_0/y_0$, increases by 45 per cent between
$r_A=0.3$ and $r_A=3 R_{\rm vir}$, while $T_X$ remains basically constant.
On the other hand, the rapid decrease of the electron temperature $T_e$ 
with $r_A$ compensates the increase of $B(\theta_b)/Y(\theta_b)$, and the
peculiar velocity estimate is basically unbiased (dashed histogram).

As for the case of spatially resolved clusters, Figure \ref{fig:vel-Te-TX-dist-apert} 
only shows the clusters with $v_{DM}> 100 $ km s$^{-1}$: slower clusters 
produce longer tails in the distributions but do not vary the central values.

The lower panels in 
Figure \ref{fig:vel-TX-Te-res-unres} show the extreme case of unresolved clusters, with
beam size $r_A=3R_{\rm vir}$. Even in this case 
the standard deviation is $\sigma=\langle (v_{\rm est}-v_{\rm DM})^2\rangle^{1/2}<160$ 
km s$^{-1}$ and therefore comparable to 
the systematic errors due to the internal bulk flows.

Figure \ref{fig:vel-Te-TX-dist-apert} clearly shows that, with only the X-ray temperature
available, it is essential to have good spatial resolution, in order to
have a measure of the ratio $B(\theta_b)/Y(\theta_b)$ 
within a sufficiently small area where $T_X$ and $T_e$ are comparable. Otherwise the large
value of $T_X$ can severely overestimate the peculiar velocity.

\section{Conclusions}\label{sec:end}

In principle, the evolution of the velocity field on large scales can be measured by
detecting the kinetic SZ effect of galaxy clusters. There exist substantial
technical difficulties in this measurement due to the weakness of the signal
and due to the fact that the kinetic effect does not produce a spectral distortion 
of the CMB.  Nevertheless, a few detections have been claimed to date
(\citealt{holz97}; \citealt{laroque03}; \citealt{benson03}; \citealt{kitayama04}), and 
more sensitive instruments will make this measurement 
less problematic in the near future.  

The estimate of the peculiar velocity of a cluster from the 
amplitude of the kinetic SZ effect assumes an isothermal ICM in hydrostatic equilibrium 
in the cluster potential well. Recent X-ray observations (e.g. \citealt{forman03}) show that the
ICM is actually far from this ideal model.

We extracted a sample of 117 fairly
realistic galaxy clusters, with mass larger than $10^{14} h^{-1} M_\odot$,  from an $N$-body hydrodynamical simulation
of a large volume of a $\Lambda$CDM cosmological model to estimate the
systematic errors, due to the gas bulk flows and ICM temperature estimate, which can affect the measures of the peculiar
velocity.

Our simulation reproduces the observed properties of X-ray clusters (\citealt{borgani03};
\citealt{ettori04}) and the distribution of the intracluster light \citep{murante04} 
reasonably well. Here, we show that our model also reproduces the observed scaling relation
between X-ray and SZ properties, namely the X-ray luminosity or the emission-weighted X-ray temperature
vs. the central value $y_0$ of the Comptonization parameter map. The results of the comparison,
between observed and simulated clusters, that we propose here, are not yet fully satisfactory,
mostly because of the lack of a uniform sample of clusters observed
both in the microwave and X-ray bands.

We then show that if the ICM is not in hydrostatic equilibrium in the cluster potential well, 
bulk flows of gas within the ICM itself can introduce a systematic error
in the estimate of the peculiar velocity. However, this effect is unlikely to be larger
than $\sim 200$ km s$^{-1}$. Our result, based on a large sample of simulated clusters, 
provides a robust statistical confirmation of the analyses of \citet{nagai03} and \citet{hold03}
who considered the $N$-body hydrodynamical simulations of only one and three clusters, respectively.

We finally investigate the relevance of the estimate of $T_e$, the ICM temperature weighted by the electron number density, 
in the peculiar velocity measurement.
A possibile systematic error originates from the fact that the observed temperature $T_X$, which is derived
from the X-ray spectrum, might be a biased estimate of $T_e$. 
In the simulations, $T_X$ is usually identified with the temperature weighted by the X-ray emission.
When averaged within large radii of the cluster centre, $T_X$
can overestimate $T_e$ by $\sim 20-50$ per cent.
If the angular resolution of the radio/microwave observation is too poor to provide
a spatially resolved profile of the SZ surface brightness and it only provides
an integrated flux, the temperature overestimate propagates into a
peculiar velocity overestimate of the same amount.

On the other hand, in the case of clusters which are spatially resolved 
in the microwave bands, we can use the temperature of the ICM in the central region of the cluster,
where $T_X$ is comparable to $T_e$. For observations which resolve a tenth
of the virial radius, the velocity is still overestimated by $10-20$ per cent,
because of the assumptions on the spatial and thermal distribution of the ICM,
usually parametrized with a $\beta$-model, used to
separate the thermal and kinetic SZ components. 

Our results rely on the usual assumption that
the emission-weighted temperature extracted from the simulations  
is comparable to the ICM temperature derived from the observed X-ray spectrum.
The emission-weighted temperature can however overestimate the spectral temperature by
$20-30$ per cent (\citealt{mathiesen01}; \citealt{mazzotta04}; \citealt{rasia04a}).
This uncertainty makes the identification of the observed $T_X$ with
$T_e$ more problematic, although it seems to alleviate the systematics
we find here. 

By combining all these results we can set a conservative lower limit of $\sim 200$ km s$^{-1}$ 
(see Figures \ref{fig:vmapscatter} and \ref{fig:vel-TX-Te-res-unres}) 
to the systematic errors that will affect upcoming measurements
of the cluster peculiar velocity with the SZ effect.
This limit is comparable to the accuracy expected when the contamination from
dusty galaxies and primary CMB anisotropies is considered in radio/microwave observations at microkelvin
sensitivity and arcminute angular resolution \citep{knox04}.
Therefore, the internal ICM bulk flows and its temperature estimate 
may limit any substantial improvements that could come from more sensitive microwave
measurements and more accurate component separations. 

We consider the systematic error we find here a lower limit because
we identify clusters in the three-dimensional space of the simulation box and analyze their X-ray and SZ properties
in two-dimensional projected maps without including instrumental noise and back-/fore-ground sources.
Real-world complications can add further systematic errors \citep{aghanim04}.
More realistic mock observations of SZ clusters are required to design a method to correct for these biases.
To accomplish this task one can construct microwave 
sky maps from $N$-body hydrodynamical simulations (e.g. \citealt{schaefer04}),
identify clusters in these maps (e.g. \citealt{melin04}) and analyze the simulated data according to standard observational procedures.
This approach will help to quantify how, in a realistic situation, the different systematic errors
propagate into the statistical analysis of cluster velocity fields. We plan to investigate this issue in future work.

\section*{ACKNOWLEDGMENTS}
We thank Gil Holder and Ian McCarthy for pointing out a few inaccuracies, 
regarding the observational data, that were present in an early version of 
the paper, and the referee, Douglas Scott, for 
relevant suggestions that improved the presentation of our results.
The simulation ran on an IBM-SP4 machine at the ``Consorzio Interuniversitario del Nord-Est per il
CAlcolo elettronico'' (CINECA, Bologna), with an INAF--CINECA CPU time grant. 
This work has been partially supported by the NATO Collaborative Linkage Grant PST.CLG.976902,
the INFN Grant PD-51, the MIUR Grant
2001, prot. 2001028932, ``Clusters and groups of galaxies: the
interplay of dark and baryonic matter'', and by ASI.
KD acknowledges support by a Marie Curie Fellowship of the European Community program
{\it Human Potential} under contract number MCFI-2001-01227.

\end{document}